\documentclass[%
 reprint,
 amsmath,amssymb,
prx,
]{revtex4-2}

\usepackage{graphicx}
\usepackage{dcolumn}
\usepackage{bm}
\usepackage[colorlinks,linkcolor=blue,anchorcolor=blue,citecolor=blue,urlcolor=blue]{hyperref}

\begin{document}

\title{Chiral spin-phonon bound states and spin-spin interactions with phononic  lattices}

\author{Xing-Liang Dong}
\author{Cai-Peng Shen}
\author{Shao-Yan Gao}
\author{Hong-Rong Li}
\author{Hong Gao}
\author{Fu-Li Li}
\author{Peng-Bo Li}
\email{lipengbo@mail.xjtu.edu.cn}
\affiliation{Ministry of Education Key Laboratory for Nonequilibrium Synthesis and Modulation of Condensed Matter, Shaanxi Province Key Laboratory of Quantum Information and Quantum Optoelectronic Devices, School of Physics, Xi'an Jiaotong University, Xi'an 710049, China}

\date{\today}

\begin{abstract}
Designing unconventional interactions between single phonons and spins
is fascinating for its applications in quantum phononics.
Here, we propose a reliable scheme for coupling spins
and phonons in phononic dimer and trimer lattices,
with the combination of solid-state defects and diamond phononic (optomechanical) crystals.
The dimer and trimer lattices used are an array of coupled phononic cavities with spatially modulated hopping rates.
We predict  a series of unconventional sound-matter interaction phenomena in this hybrid quantum system.
In the dimer lattice, we show the formation of chiral spin-phonon bound states
and topology-dependent phononic collective radiation.
While in the trimer lattice, chiral bound states still exist and
the spin relaxation is sublattice-dependent.
The chiral bound states existed in both types of lattices are robust to large amount of disorder,
which can mediate chiral and robust spin-spin interactions.
This work provides a promising platform for  phonon-based quantum information processing and quantum simulation.

\end{abstract}

\maketitle


\section{introduction}

Engineering and controlling dipole-dipole interactions is at the heart of analog quantum simulation,
which can be realized through manipulating the field environment between emitters \cite{RevModPhys.86.153}.
A fascinating platform is the nanophotonic structure, wherein the dispersion relation can be designed on-demand,
and the atom-photon bound states can form within the band and band-gap when atoms are placed nearby
\cite{PhysRevLett.64.2418,Subwavelength2015AGT,Quantum2015Douglas,Hood10507,PhysRevA.97.043831,PhysRevLett.125.163602}.
The photonic component in the bound state is virtually excited, thereby can mediate long-range tunable dipole-dipole interactions without dissipations into the guided modes.
Usually, the photon bound state is exponentially localized and isotropic around the emitter, which limits the form of dipole-dipole interactions
\cite{Subwavelength2015AGT,Quantum2015Douglas}.
However, through properly modifying the photonic bath, photon bound states can be anisotropic
\cite{Belloeaaw0297,PhysRevX.11.011015,doi:10.1021/acsphotonics.8b01455,PhysRevLett.126.043602},
power-law scaling \cite{PhysRevA.97.043831,PhysRevLett.125.163602,PhysRevA.103.033511}, and even phase tunable
\cite{PhysRevResearch.2.023003,PhysRevLett.126.063601,PhysRevLett.126.103603,PhysRevLett.126.203601}.
These special bound states can be used to study more exotic many-body phases.
In addition, there are other new quantum phenomena when emitters interact with structured photonic baths,
which have potential applications in quantum information processing
\cite{PhysRevLett.115.063601,PhysRevLett.119.143602,PhysRevA.96.043811,Barik666,PhysRevLett.122.203603,PhysRevA.99.053852}.

A new paradigm to capture novel bound states is taking advantage of topological photonic lattices
\cite{PhysRevA.97.043831,Belloeaaw0297,PhysRevX.11.011015,PhysRevLett.125.163602,
PhysRevA.103.033511,PhysRevLett.126.063601,vega2021qubitphoton}.
In particular, in the dimer photonic lattice (a photonic analog of the Su-Schrieffer-Heeger (SSH) model),
there exist chiral photonic bound states with only one side envelope with respect to the emitter
\cite{Belloeaaw0297,PhysRevX.11.011015}.
These directional bound states can mediate directional dipole-dipole interactions.
As a natural extension, the trimer lattice allows for  chiral edge states that appear at one side of the lattice \cite{PhysRevA.99.013833}.
These edge states have a direct connection with that of two-dimensional Aubry-Andr\'{e}-Harper (AAH) models
\cite{PhysRevA.96.032103},
thus is robust against disorders.
Using such chiral edge modes, the adiabatic topological pumping
has been predicted and experimentally observed in 1D phononic lattices
\cite{PhysRevLett.123.034301,PhysRevLett.126.095501}.
However, so far the interaction with emitters and emitter-emitter interactions in this type of photonic or phononic lattice have been unexplored.

On the other hand, phonon, the quanta of mechanical vibration, is regarded as an alternative quantum information carrier
\cite{Habraken_2012,Gustafsson207,PhysRevX.5.031031,PhysRevLett.117.015502,Chu199,PhysRevLett.120.213603,
Quantum2018Satzinger,Bienfait368,PhysRevLett.124.053601,PhysRevLett.125.153602,doi:10.1063/5.0024001,phononic2021Neuman}.
The spin-mechanical hybrid systems may overcome the shortcoming of vacuum radiations in nanophotonic structures,
because  phonons do not decay into free-space.
One of the most promising platforms is diamond crystal
\cite{Burek:16,Sipahigil847,Evans662,CHIA2021219}, since in terms of nanofabrication,
it is capable of fabricating high-quality mechanical modes at the GHz frequency,
while integrating color centers as long-lived spin qubits
\cite{Balasubramanian2009Ultralong,Maurer1283,Bar2013Solid,PhysRevLett.112.036405,Tao2014Single,
Lee2017Topical,PhysRevLett.118.223603,PhysRevLett.119.223602,Quantum2019Bradac,Coupling2021Kazuhiro,Quantum2021Romain}.
So far, the strain coupling mechanism between diamond electric spins and mechanical modes has been widely explored,
providing new opportunities for realizing phonon networks in the strong-coupling regime
\cite{PhysRevLett.110.156402,PhysRevB.88.064105,PhysRevLett.111.227602,PhysRevLett.113.020503,
Dynamic2014Ovartchaiyapong,Strong2015Barfuss,PhysRevLett.116.143602,PhysRevX.6.041060,Controlling2018Sohn,
PhysRevB.97.205444,PhysRevX.8.041027,Coherent2020Maity,PhysRevA.101.042313,PhysRevA.103.013709}.
In particular, diamond phononic crystal waveguides can host phononic band gaps and the spin-phonon bound states can form
when the spin's frequency lies within this special frequency range
\cite{BandPeng,PhysRevResearch.2.013121,PhysRevResearch.3.013025}.
Such spin-phonon bound states can mediate long-range tunable spin-spin interactions for simulating spin models  in hybrid quantum systems.
Moreover, the phononic topological states as well as their interactions with one or more emitters
at the quantum level are widely investigated in phononic crystals
\cite{PhysRevX.5.031011,Lemonde_2019,PhysRevB.101.085108,https://doi.org/10.1002/adfm.201904784}.
Combining topological phononic structures with solid-state spins may give rise to exotic spin-phonon bound states
and other interesting sound-matter interaction phenomena.

In this work, we consider  sound-matter interactions  in dimer and trimer phononic lattices,
where an array of solid-state defects are integrated into a diamond phononic crystal.
The dimer and trimer lattices used are an array of coupled phononic crystal cavities with spatially modulated hopping rates.
In contrast to previous works where the phononic edge modes are obtained from the breakdown of the time-reversal symmetry
with the assistance of optical modes \cite{PhysRevX.5.041002,Lemonde_2019,Two2020Ren},
here we obtain the edge states through the periodic modification of the hopping rates in the 1D phononic crystal.

In the dimer lattice, we show the formation of chiral spin-phonon bound states,
where the spin mimics the behaviour of a boundary to localize phonons on its one side.
Furthermore, when two spins are considered, the collective decay depends on the phononic waveguide structure between them,
and more interestingly, the phonon collective decay can be topology-dependent.
While in the trimer lattice,
the chiral bound states still exist with a number of six that corresponds to six chiral edge states,
occurring at certain energy and sublattices.
When the spin's frequency is resonant with the band, we show the spin relaxation becomes sublattice-dependent as a result of mirror symmetry breaking.
Furthermore, due to the topology or topological origin, the chiral spin-phonon bound states in both lattice structures
are robust to large amounts of disorder, which can mediate chiral and robust spin-spin interactions.
These interactions enable the exploration of exotic many-body phases such as double N\'{e}el ordered states
\cite{Belloeaaw0297,bello2021spin}.
The quantum control over phononic edge states is also considered in a finite system
in terms of population inversion between spins and edge states.
We also illustrate how to design the required lattice structure in a 1D diamond nanobeam
and show that our scheme is feasible under current experimental conditions.
This work provides a novel platform for manipulating phonons and is useful for phonon-based quantum applications.

\section{The setup}

\begin{figure}
\includegraphics[scale=0.14]{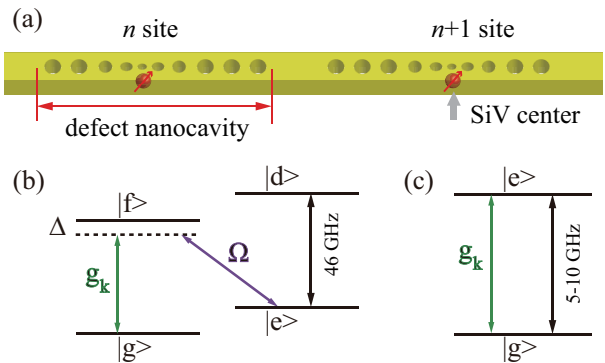}
\caption{\label{fig1}(Color online) (a) Schematic of an array of phononic cavites in a diamond nanobeam
with integrated SiV centers. (b,c) Level structure of electronic ground state of single SiV centers
and strain coupling of SiV spins to mechanical modes via a Raman process (b) or directly (c).}
\end{figure}

We consider a full lattice of phononic cavities in diamond crystals,
where a single solid-state spin is integrated into each lattice site, as depicted in Fig.~1(a).
The acoustic resonators have identical resonance frequency $\omega_m$ and
each pair of adjacent ones are coupled via either direct near-field coupling or a phonon waveguide,
wherein the hopping rates are tunable via structure designs.
The Hamiltonian of this coupled phononic cavity waveguide is written as (setting $\hbar=1$)
\begin{eqnarray}\label{ME1}
\hat{H}_\text{ph}=\omega_m\sum_{n}\hat{d}_n^\dag\hat{d}_n-\sum_{n}(J_n\hat{d}_n^\dag\hat{d}_{n+1}+\text{H.c.}),
\end{eqnarray}
where $\hat{d}_n$ is the annihilation operator of the phononic mode,
and $J_n$ is the tunnelling strength between the $n$th cavity and the $(n+1)$th cavity.

The solid-state spins considered in this setup are SiV centers,
whose electric ground states have a splitting of $\sim46$ GHz
and possess a large strain susceptibility due to the spin-orbit coupling.
In the presence of a static magnetic field, the two orbit states are split into four energy levels, as shown in Fig.~1(b).
The sublevels $|g\rangle$ and $|e\rangle$ can be viewed as a long-lived spin,
which can be  indirectly coupled to the acoustic modes with frequencies of  $46$ GHz
via a Raman process (see Fig.~1(b)) or directly coupled to the acoustic modes with frequencies of several GHz (see Fig.~1(c))
\cite{PhysRevLett.120.213603,PhysRevB.97.205444}.
In both cases, the single phonon coupling strength can reach MHz such that the strong coupling condition can be satisfied.
Under the rotating wave approximation, the total Hamiltonian of the system is
\begin{eqnarray}\label{ME2}
\hat{H}_\text{tot}=\hat{H}_\text{ph}+\sum_{n}\omega_\sigma^{(n)}\hat{\sigma}_+^n\hat{\sigma}_-^n
+g\sum_{n}(\hat{\sigma}_{+}^n\hat{d}_n+\text{H.c.}),
\end{eqnarray}
with $\omega_\sigma^{(n)}$ the (effective) resonance frequency of the $n$th spin,
$\hat{\sigma}_+^n=|e\rangle^n\langle g|$,
and $g$ the (effective) spin-phonon coupling strength.

\section{Spin-phonon interactions in a dimer lattice}

\subsection{The Hamiltonian}

\begin{figure}
\includegraphics[scale=0.18]{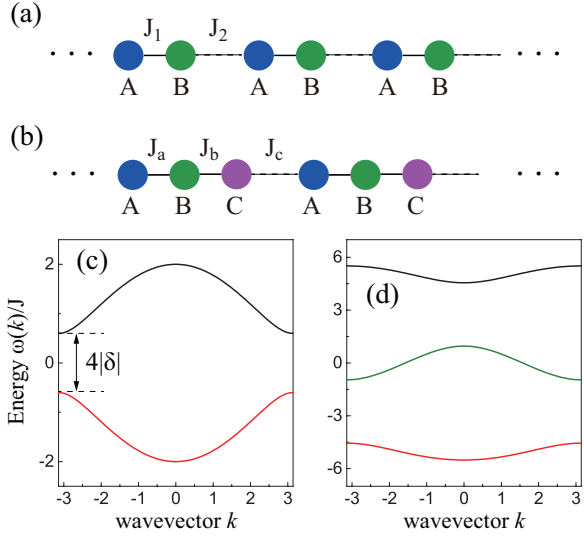}
\caption{\label{fig1}(Color online) Schematic diagram of dimer (a) and trimer (b) lattice.
(c) Dispersion relations of the model in (a), with $J_1=J(1+\delta)$, $J_2=J(1-\delta)$ and $\delta=\pm0.3$.
(d) Dispersion relations of the model in (b), with $(J_a,J_b,J_c)=(1,4,3)J$.}
\end{figure}

When the hopping rates are spatially designed with a periodicity of $2$, the phononic crystal behaves as a dimer lattice,
which is the phononic analog of the Su-Schrieffer-Heeger (SSH) model
and allows for topological phonons at the quantum level.
The Hamiltonian in Eq.~(\ref{ME1}) can be rewritten as
\begin{eqnarray}\label{ME3}
\hat{H}_\text{ph}&=&\omega_m\sum_{n}(\hat{a}_n^\dag\hat{a}_n+\hat{b}_n^\dag\hat{b}_n)\notag\\
&-&\sum_{n}(J_1\hat{a}_n^\dag\hat{b}_n+J_2\hat{b}_n^\dag\hat{a}_{n+1}+\text{H.c.}),
\end{eqnarray}
where $\hat{a}_n$ and $\hat{b}_n$ are respectively the annihilation operator of phonon
at sublattice $A$ and sublattice $B$ of the $n$th unit cell.
The hopping rates can be replaced by $J_1=J(1+\delta)$, and $J_2=J(1-\delta)$,
with $\delta$ the dimerization strength.
To transform the Hamiltonian into momentum space, we
impose periodic boundary conditions and introduce the Fourier transformation
\begin{eqnarray}\label{ME4}
\hat{d}_k=\frac{1}{\sqrt{N}}\sum_{n=1}^{N}e^{-iknd_0}\hat{d}_n.
\end{eqnarray}
For simplicity, we take the lattice constant $d_0=1$ below.
Then the Hamiltonian of the phononic bath is transformed into
$\hat{H}_{\text{ph}}=\sum_k\hat{V}_k^\dag\hat{H}(k)\hat{V}_k$,
with $\hat{V}_k^\dag=(\hat{a}_k^\dag,\hat{b}_k^\dag)$ and the kernel
\begin{eqnarray}\label{ME5}
\hat{H}(k)=
{\left( \begin{array}{ccc}
\omega_m & -J_1-J_2e^{-ik}\\
-J_1-J_2e^{ik} & \omega_m
\end{array}\right )}.
\end{eqnarray}
Diagonalizing the Hamiltonian and taking $\omega_m$ as the energy reference,
we can get the eigenenergies, that is, the dispersion relation
\begin{eqnarray}\label{ME6}
\omega_{\alpha/\beta}(k)&=&\pm\sqrt{J_1^2+J_2^2+2J_1J_2\cos k}\notag\\
&=&\pm J\sqrt{2(1+\delta^2)+2(1-\delta^2)\cos k}.
\end{eqnarray}
The two dispersions are symmetric with respect to the zero energy due to chiral symmetry and we set $\omega_{\alpha}(k)=-\omega_{\beta}(k)\equiv\omega(k)$ below.
The band structure is shown in Fig.~2(c), where a middle bandgap with size $4J|\delta|$ is opened.
In particular, for $\delta<0$, it corresponds to a topological phase of sound
and there exists a pair of phononic edge states exponentially localized at the two ends when the system is finite
\cite{asboth2016short}.

We now consider the spins are coupled to the topological phonons in dimer lattices.
The interaction Hamiltonian of the spins and bath in $k$-space is given by
\begin{eqnarray}\label{ME7}
\hat{H}_{\text{int}}&=&\frac{g}{\sqrt{2N}}\sum_{k,m=A}e^{ikx_m}\hat{\sigma}_{+}^m(\hat{\alpha}_k-\hat{\beta}_k)\notag\\
&+&\frac{g}{\sqrt{2N}}\sum_{k,m=B}e^{i(kx_m-\phi(k))}\hat{\sigma}_{+}^m(\hat{\alpha}_k+\hat{\beta}_k)+\text{H.c.},
\end{eqnarray}
with $\hat{\alpha}_k$ and $\hat{\beta}_k$ the eigenoperators in the diagonalization basis, $x_m$ the unit cell position
and $\phi(k)\equiv\text{arg}(-J_1-J_2e^{-ik})$.
The first (second) term represents the interactions between the phonons and spins located at the $A$ ($B$) sublattice.
The single spins can acquire extra energy due to their interaction with the quantized phononic field,
which is the self-energy expressed as
\begin{eqnarray}\label{ME8}
\Sigma_e(z)=\sum_{k,i=j}\sum_{s=\alpha,\beta}\frac{\langle 0|\hat\sigma_-^j\hat{H}_\text{int}\hat{s}_k^\dag|0\rangle\langle 0|\hat{s}_k\hat{H}_\text{int}\hat\sigma_+^i|0\rangle}{z-\omega_s(k)},
\end{eqnarray}
with $|0\rangle=|g\rangle|\text{vac}\rangle$.
In particular, within the Markovian approximation,
$-2\text{Im}\Sigma_e(z)$ describes the decay rate and $\text{Re}\Sigma_e(z)$ represents the energy shift.
The interaction with the phononic bath can also lead to collective self-energy $\Sigma_{ij}(z)$ of the spins,
which can be obtained by simply replacing $i\neq j$ in the summation of $\Sigma_e$.
Since the phononic dimer lattice has sublattice (chiral) symmetry,
the pairwise collective self-energy can be classified into two categories:
one  from the spins in the same sublattice and the other  from the spins in the different sublattice
\begin{eqnarray}\label{ME9}
&\Sigma&_{ij}^{AA/BB}(z)=\pm\frac{-g^2zy_\text{min}^{|x_{ij}|}}{\sqrt{z^4-4J^2(1+\delta^2)z^2+16J^4\delta^2}}\\
&\Sigma&_{ij}^{AB}(z)=\pm\frac{g^2J[(1+\delta)y_\text{min}^{|x_{ij}|}+(1-\delta)y_\text{min}^{|x_{ij}+1|}]}
{\sqrt{z^4-4J^2(1+\delta^2)z^2+16J^4\delta^2}}\label{ME10},
\end{eqnarray}
where $x_{ij}=x_j-x_i$ is the cell distance between two spins,
$y_\text{min}=min(y_+,y_-)$ with respect to their absolute values and,
\begin{eqnarray}\label{ME11}
y_\pm=\frac{z^2-2J^2(1+\delta^2)\pm\sqrt{z^4-4J^2(1+\delta^2)z^2+16J^4\delta^2}}{2J^2(1-\delta^2)}.
\end{eqnarray}
The sign in these two self-energy expressions is positive when $y_\text{min}=y_+$ and negative when $y_\text{min}=y_-$.
The self-energy $\Sigma_e(z)=\Sigma_{i=j}^{AA/BB}(z)$ of single spins is also given here.

\subsection{Chiral spin-phonon bound states}

\begin{figure}
\includegraphics[scale=0.18]{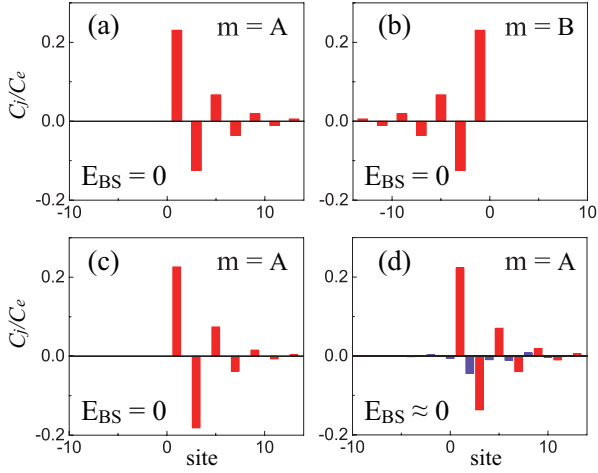}
\caption{\label{fig1}(Color online) Chiral spin-phonon bound states at zero energy.
(a,c,d) Single spins are coupled to the $A$ sublattice. (b) Single spins are coupled to the $B$ sublattice.
(c) With off-diagonal disorders. (d) In the presence of on-site disorders.
The disorder strength in both cases is $\xi_n\in[-J/4,J/4]$.
Note that the spin-phonon coupling strength is $g=0.3J$.}
\end{figure}

When the quantum emitters are coupled to this bath,
the related topological features will be imprinted into the emitter-bath interaction.
Specially, if the emitters' frequency is chosen to be strictly equal to the zero-energy modes,
the emitters can act as an effective edge of the lattice and localize the excitation on only one side
\cite{Belloeaaw0297,PhysRevX.11.011015}.
In this section, we show the formation of chiral spin-phonon bound states in this SiV-phononic crystal model.

We consider single spins coupled to the phononic dimer lattice, whose frequency lies within the bandgap.
The spin no longer decays into the propagating modes, but is dressed by the acoustic modes close to the bandedges.
The stationary state wavefunction in the single-excitation subspace can be expanded by
\begin{eqnarray}\label{ME12}
|\psi\rangle=(C_e\hat{\sigma}_{+}+\sum_{k}\sum_{s=a,b}C_{k,s}\hat{s}_k^\dag)|g\rangle|\text{vac}\rangle,
\end{eqnarray}
with $C_e$ the probability amplitude of the spin being in the upper state and
$C_{k,s}$ the probability amplitude for finding a phononic excitation in sublattice $A/B$ at the wavevector $k$.
The coefficients can be obtained by solving the secular equation
$\hat{H}_\text{tot}|\psi\rangle=E_\text{BS}|\psi\rangle$, yielding
\begin{eqnarray}\label{ME13}
C_{k,a}^{A(B)}&=&\frac{gC_eE_\text{BS}}{E_\text{BS}^2-\omega^2(k)}\quad \bigg(\frac{gC_e\omega(k)e^{i\phi(k)}}{E_\text{BS}^2-\omega^2(k)}\bigg)\\
C_{k,b}^{A(B)}&=&\frac{gC_e\omega(k)e^{-i\phi(k)}}{E_\text{BS}^2-\omega^2(k)}\quad
\bigg(\frac{gC_eE_\text{BS}}{E_\text{BS}^2-\omega^2(k)}\bigg)\label{ME14},
\end{eqnarray}
where the superscript in the wavefunction labels the sublattice (at the $j=0$ unit cell) to which  the spin is coupled.
Doing the Fourier transformation, we can obtain the corresponding spatial distribution in sublattice $A/B$.

In this work, we mainly focus on the bound states at zero energy.
A direct observation of the expressions shows that, when setting $E_\text{BS}=0$,
$C_{j,a}^{A}=C_{j,b}^{B}=0$, with $j$ the unit cell index.
The other two spatial distributions can be given analytically, similar to
the calculation of the self-energy.
For the case of $\delta>0$ and the spin at the $j=0$ unit cell,
\begin{eqnarray}\label{ME15}
C_{j,b}^{A}=\left\{
\begin{aligned}
&\frac{gC_e(-1)^{j}}{J(1+\delta)}\bigg(\frac{1-\delta}{1+\delta}\bigg)^{j}, &j\geq0\\
&0, &j<0
\end{aligned}
\right.
\end{eqnarray}
and $C_{j,a}^{B}=C_{-j,b}^{A}$,
which indicates that the left (right) bound state vanishes when the spin is coupled to $A$\,($B$) sublattice.
The chirality of the bound states arises from the inversion symmetry breaking of the phononic bath
with respect to the coupling point, and reaches its maximum at $\omega_\sigma=\omega_m$.
These two perfect chiral bound states are shown in Fig.~3(a) and 3(b),
with numerical integrations of Eq.~(\ref{ME13}) and Eq.~(\ref{ME14}).
Note that the number of chiral bound states is equal to that of the edge states.

We next consider the robustness of this chiral spin-phonon bound states in the dimer lattice.
A finite chain with $40$ phononic cavities is used to simulate the SSH bath.
The chiral bound states in the presence of disorder in the phonon hopping strength are depicted in Fig.~3(c),
with adding the random terms $\hat{H}_d=\sum_{n}(\xi_{n}\hat{d}_n^\dag\hat{d}_{n+1}+\text{H.c.})$ to the total Hamiltonian.
The disorder strength $\xi_{n}$ is chosen within the range $[-W/2, W/2]$ for the $n$th lattice site.
We show that the chiral behavior of the bound states is protected, where the chiral symmetry persists.
As a contrast, the diagonal disorder is also taken into account, which is shown in Fig.~3(d).
We find that the middle bound state now has distributions in the A/B sublattices and both sides of the spin.
Besides, the energy of the bound state is no longer strictly fixed at zero.

\subsection{Topology-dependent  collective radiation}

\begin{figure}
\includegraphics[scale=0.24]{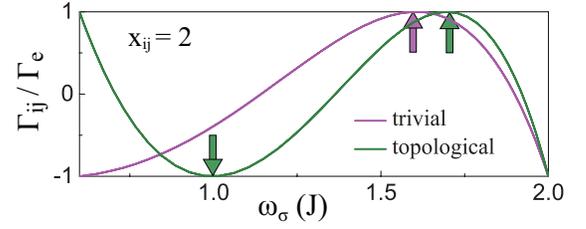}
\caption{\label{fig1}(Color online)
The collective decay rate $\Gamma_{ij}$ of two spins, normalized to $\Gamma_e$, versus the spin's frequency in the upper band.
Here, the distance between the spins is fixed with $x_{ij}=2$,
$\delta=0.3$ for the trivial phase (purple line) and $\delta=-0.3$ for the topological phase (green line).}
\end{figure}

Though the main topological feature is most pronounced in the zero-energy modes,
the dynamics of spins in the phononic dimer lattice is different from that in the standard waveguide.
In the weak coupling limit, the spin dynamics is subject to Eq.~(\ref{ME9}) and Eq.~(\ref{ME10}),
therein the imaginary part represents the decay rate.
From Eq.~(\ref{ME9}) (in the case $i=j$), we show that the decay rate of the spins is sublattice-independent,
which can be roughly understood from the viewpoint that the spins couple to the two topology-different semi-infinite waveguides.
Without loss of generality, we consider two spins resonant with the phononic band.
There are three cases: i) the spins are coupled to $AA/BB$ sublattice;
ii) the spins are coupled to $AB$ sublattice (from left to right);
iii) the spins are coupled to $BA$ sublattice.
In these three situations, the waveguide structures between the spins
are standard-like, trivial and topological, respectively.
As a result, the phonon-induced collective decay rate is different for the three cases.
More interestingly, case II and case III are interchangeable
by only varying the sign of the parameter $\delta$.
Therefore, the collective radiation is topology-dependent.

The collective decay rate $-2\text{Im}\Sigma_{ij}$ can also be calculated by using
the identical relation $\lim_{y\rightarrow0^+}1/(x\pm iy)=P(1/x)\mp i\pi\delta(x)$,
which can provide a more intuitive insight to the features above.
In the standard waveguide, the collective decay rate oscillates versus the emitter's relative position, i.e.,
$\Gamma_{ij}=\Gamma_e\cos(kx_{ij})$.
In contrast, the collective decay rate in the SSH bath is given by
\begin{eqnarray}\label{ME16}
\Gamma_{ij}^\text{AA/BB}(\omega)&=&\text{sign}(\omega)\Gamma_e(\omega)\cos(k(\omega)x_{ij})\\
\Gamma_{ij}^\text{AB/BA}(\omega)&=&\text{sign}(\omega)\Gamma_e(\omega)\cos(k(\omega)x_{ij}\mp\phi(k))\label{ME17}.
\end{eqnarray}
Here, $x_{ij}$ is the cell-distance.
Obviously, the topology-dependent parameter (phase) $\phi(k)$ enters into Eq.~(\ref{ME17}).
In Fig.~4, we plot $\Gamma_{ij}^\text{AB}(\omega)$ as a function of spin frequency $\omega_\sigma$ inside the upper band,
with setting $x_{ij}=2$ and $\delta=\pm0.3$.
The case is opposite for $\omega_\sigma$ inside the lower band, i.e.,
$\Gamma_{ij}^\text{AB}(\omega)=-\Gamma_{ij}^\text{AB}(-\omega)$.
Perfect super or subradiance occurs when $\Gamma_{ij}^\text{AB}=\pm\Gamma_{e}$.
In particular, we find that the number of phononic super or subradiant points depends on the sign of $\delta$:
In the trivial phase ($\delta>0$), there are $x_{ij}-1$ perfect superradiant points inside the two passbands,
while in the topological phase ($\delta<0$), there are $x_{ij}$ frequency points.

\subsection{Chiral spin-spin interactions}

\begin{figure}
\includegraphics[scale=0.18]{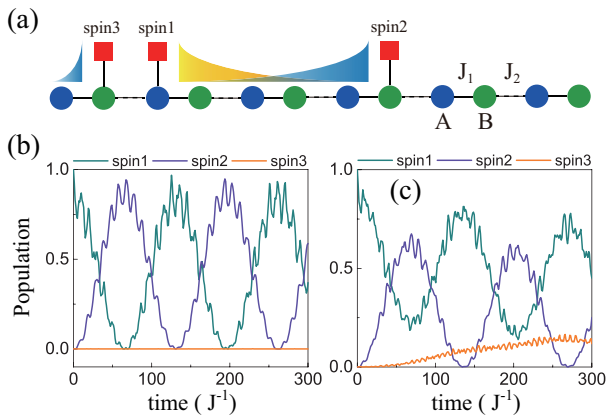}
\caption{\label{fig1}(Color online)
Chiral and robust spin-spin interactions in the trivial phase ($\delta=0.3$) of the dimer lattice.
(a) Three spins are respectively coupled to the second, third and $8$th phononic cavity of a 12 site cavity chain.
Chiral spin-phonon bound states are plot.
(b,c) Population of three spins with time evolution, with the initial state $|\psi_0\rangle=|1\rangle_1|0\rangle_2|0\rangle_3$.
(b) With disorder on the hopping rates $J_1$ and $J_2$. (c) With diagonal disorder.
The disorder strength in both cases is $\xi_n\in[-J/2,J/2]$.}
\end{figure}

The  chiral spin-phonon bound states discussed above can mediate spin-spin interactions
through the exchange of virtual phonons.
As a result, the spin-spin interactions will inherit the features appearing in the bound states:
i) decay exponentially with the spins' distance;
ii) the spin located in sublattice $A$\,($B$) only interacts with spin located in sublattice $B$\,($A$) on the one side;
iii) robust against the off-diagonal disorder.
For $\delta>0$, the interaction in the Markovion limit is given as
\begin{eqnarray}\label{ME18}
\hat{H}_{s-s}=\sum_{i<j}J_{ij}(\hat{\sigma}_+^i\hat{\sigma}_-^j+\hat{\sigma}_+^j\hat{\sigma}_-^i)
\end{eqnarray}
with the coupling strength $J_{ij}^{AA/BB}=0$ and
\begin{eqnarray}\label{ME19}
J_{ij}^{AB}=\left\{
\begin{aligned}
&\frac{g^2(-1)^{x_{ij}}}{J(1+\delta)}\bigg(\frac{1-\delta}{1+\delta}\bigg)^{x_{ij}}, &x_{ij}\geq0\\
&0, &x_{ij}<0
\end{aligned}
\right..
\end{eqnarray}

To numerically show these chiral spin-spin interactions, we use a finite lattice chain with $12$ phononic cavities
and consider three spins coupled to the second, third and $8$th cavity (see Fig.~5(a)).
The off-diagonal disorder is also taken into account.
The quantum dynamics of the three spins is plot in Fig.~5(b),
with the initial state $|\psi_0\rangle=|1\rangle_1|0\rangle_2|0\rangle_3$.
We find that the population of spin 3 is always zero and the population is oscillating between
spin 1 and spin 2 with an amplitude approaching one,
which reflects the directionality and robustness of the spin-spin interactions.
Similarly, the quantum dynamics of spins in the presence of diagonal disorder is plot in Fig.~5(c) for comparison.
Apart from extra population in spin 3, we find that the population of spin 1 can not reach zero, since
the energy value of the bound state is no longer pinned at zero in this case.

\subsection{Quantum control over phononic edge states}

\begin{figure}
\includegraphics[scale=0.23]{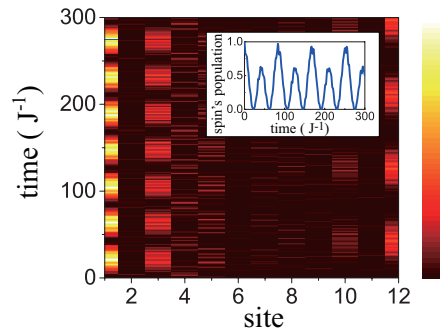}
\caption{\label{fig1}(Color online)
Control of phononic edge states in dimer lattice
in a finite system composed of $12$ phononic cavities, with a single spin inside the $5$th cavity.
The time evolution of population in cavities array and the spin is plot. Here, we choose $\delta=-0.3$.}
\end{figure}

In the finite system, edge effects inevitably occur especially in the topological phase,
where a pair of edge states exponentially localize at the two ends of the lattice chain.
The phononic edge states in a phononic crystal can be controlled with the addition of solid-state spins.
In that case, the excitation can be transferred between spins and phononic edge states,
with eigenenergies $\pm\epsilon$ and eigenstates $|E_\pm\rangle$.
The dynamics is govern by the effective Hamiltonian
\begin{eqnarray}\label{ME20}
\hat{H}_\text{eff}&=&\sum_{\pm}\pm\epsilon|E_\pm\rangle\langle E_\pm|+(\omega_\sigma-\omega_m)|e\rangle\langle e|\notag\\
&+&\sum_{\pm}g_\pm(|e\rangle\langle E_\pm|+|E_\pm\rangle\langle e|),
\end{eqnarray}
with $g_\pm=g\langle E_\pm|\hat{d}_{x_0}^\dag|\text{vac}\rangle$
and $g_+=g_-$ ($g_+=-g_-$) when $\hat{d}_{x_0}^\dag=\hat{b}_{x_0}^\dag$ ($\hat{d}_{x_0}^\dag=\hat{a}_{x_0}^\dag$).
When setting $\omega_\sigma=\omega_m$ and applying the Schr\"{o}dinger equation,
we can obtain the coupled equations of the coefficients
\begin{eqnarray}\label{ME21}
i\dot{C}_e(t)&=&g_+C_+(t)+g_-C_-(t)\\
i\dot{C}_+(t)&=&\epsilon C_+(t)+g_+C_e(t)\\\label{ME22}
i\dot{C}_-(t)&=&-\epsilon C_-(t)+g_-C_e(t)\label{ME23},
\end{eqnarray}
with $C_e$ and $C_\pm$ the wavefuncions in the states $|e\rangle$ and $|E_\pm\rangle$, respectively.
Solving these equations, the spin's time evolution obeys
\begin{eqnarray}\label{ME24}
C_e(t)=\frac{\epsilon^2+2g_+^2\cos(\omega_0t)}{\epsilon^2+2g_+^2},
\end{eqnarray}
with $\omega_0=\sqrt{\epsilon^2+2g_+^2}$.
In the case of a large size cavity chain, $\epsilon\simeq0$ and
the spin exchanges energies between the left or right phononic edge state,
which are symmetry and antisymmetry superpositions of the states $|E_\pm\rangle$, respectively.

Concretely, we consider a single spin coupled to the $5$th phononic cavity of
a $12$ cavity chain designed with staggered hopping strengths and being in the topological phase.
The system's dynamics is shown in Fig.~6.
We show that the excitation is mainly transferred between the spin and the left phononic edge state,
in line with the prediction based on Eq.~(\ref{ME24}).

\section{Spin-phonon interactions in trimer lattices}

Previous work on trimer lattices have predicted the emergence of chiral edge states localized at one end of the system,
without a counterpart on the opposite edge at the same energy in the phase without inversion symmetry protection
\cite{PhysRevA.99.013833,PhysRevLett.123.034301,PhysRevLett.126.095501}.
Though there is no topology (no symmetry), these chiral edge states are still robust
against large amounts of disorder due to the topological origin.
Therefore, it's still possible to  obtain chiral spin-phonon bound states where the spin acts as the effective edges, and
 realize directional and robust quantum state transfer as well as quantum control over phononic edge states in a trimer lattice.
In addition, the emission dynamics of the spins in this lattice is worth investigating
since it is different from the one in a symmetry protected system.

\subsection{Hamiltonian}

Similar to the dimer lattice, a trimer lattice is modeled with alternating hopping rates and
having a periodicity of 3. The Hamiltonian of the phononic waveguide is rewritten as
\begin{eqnarray}\label{ME25}
\hat{H}_\text{ph}&=&\omega_m\sum_{n}(\hat{a}_n^\dag\hat{a}_n+\hat{b}_n^\dag\hat{b}_n+\hat{c}_n^\dag\hat{c}_n)\notag\\
&-&\sum_{n}(J_a\hat{a}_n^\dag\hat{b}_n+J_b\hat{b}_n^\dag\hat{c}_{n}+J_c\hat{c}_n^\dag\hat{a}_{n+1}+\text{H.c.}),
\end{eqnarray}
with $\hat{a}_n$, $\hat{b}_n$ and $\hat{c}_n$ the annihilation operators
for the $A$, $B$ and $C$ phononic modes at the $n$th unit cell.
In a common trimer lattice, the phonon tunneling strengths $J_a$, $J_b$ and $J_c$ are unequal and there is no symmetry.
The inversion symmetry only recovers in the special case of $J_a=J_b\neq J_c$.
Performing the same operation as in Sec.~III.A, the Hamiltonian can be transformed into momentum space,
with the kernel of Hamiltonian
\begin{eqnarray}\label{ME26}
\hat{H}(k)=
{\left( \begin{array}{ccc}
\omega_m & -J_a & -J_ce^{-ik}\\
-J_a & \omega_m & -J_b\\
-J_ce^{ik} & -J_b & \omega_m
\end{array}\right )}.
\end{eqnarray}
Unlike  the dimer lattice where  the analytical solution to the dispersion relation is given,
here we introduce a unitary matrix $M=(V_1,V_2,V_3)$ to diagonalize the kernel of Hamiltonian as
$\hat{D}(k)=M^\dag\hat{H}(k)M=\text{diag}(\omega_\alpha,\omega_\beta,\omega_\gamma)$,
with $V_i$ ($i\!=\!1,2,3$) the eigenvectors and $\omega_s$ ($s\!=\!\alpha,\beta,\gamma$) the eigenvalues.
The eigenoperators are given by
$(\hat\alpha_k\,\,\hat\beta_k\,\,\hat\gamma_k)^T=M^\dag(\hat{a}_k\,\,\hat{b}_k\,\,\hat{c}_k)^T$.
Then the Hamiltonian of the phononic trimer lattice in $k$-space reads
\begin{eqnarray}\label{ME27}
\hat{H}_\text{ph}=\sum_{k}\sum_{s=\alpha,\beta,\gamma}\omega_s(k)\hat{s}_k^\dag\hat{s}_k.
\end{eqnarray}
The band structure is numerically plotted in Fig.~2(d), with $J_a=J$, $J_b=4J$ and $J_c=3J$.

Working in $k$-space,
the interaction Hamiltonian for the spins and phonon modes in the trimer lattice becomes
\begin{eqnarray}\label{ME28}
\hat{H}_{\text{int}}=\frac{g}{\sqrt{N}}\sum_{k}\sum_{m,s}
e^{ikx_m}M(m,s)\hat{\sigma}_{+}^m\hat{s}_k+\text{H.c.},
\end{eqnarray}
where the interaction is projected to the new basis.
The self-energy of the spins is expressed as
\begin{eqnarray}\label{ME29}
\Sigma_e(z)=\frac{g^2}{2\pi}\int_{-\pi}^{\pi}dk\sum_{m=n,s}
\frac{\hat{M}(n,s)\hat{M}^{-1}(s,m)}{z-\omega_s(k)}.
\end{eqnarray}
The collective self-energy $\Sigma_{ij}$ of the spins can be obtained by
simply replacing $m\neq n$ in the summation of $\Sigma_e$
and adding a propagation factor $e^{ik(x_n-x_m)}$ to the integral.

\begin{figure}
\includegraphics[scale=0.13]{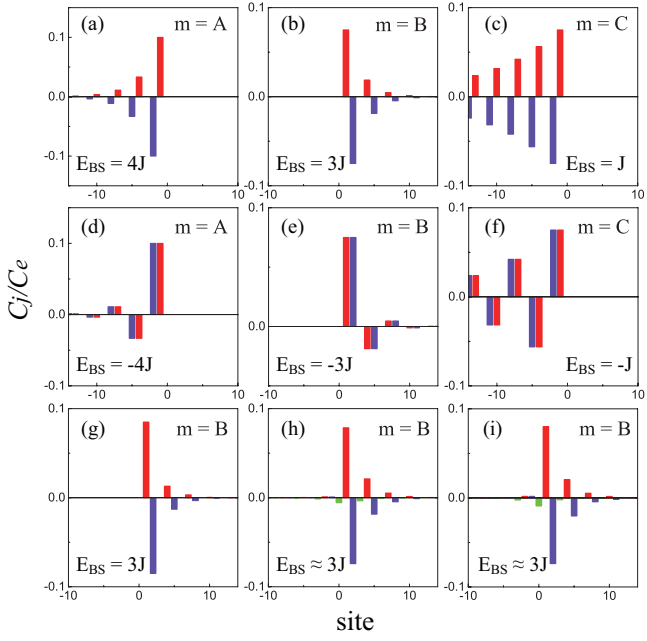}
\caption{\label{fig1}(Color online)
Chiral spin-phonon bound states in a trimer lattice, with $(J_a,J_b,J_c)=(1,4,3)J$.
(a-f) Without disorder. (g) Disorder in the intracell hopping $J_a$ and $J_b$.
(h) Disorder in the intercell hopping $J_c$. (i) Disorder in the phononic frequency $\omega_m$.
The disorder strength in the three cases is $\xi_n\in[-J/2,J/2]$.}
\end{figure}

\subsection{Chiral spin-phonon bound states}

When the spin's frequency lies within the phononic bandgap, there exist
spin-phonon bound states in the single-excitation subspace.
The spatial distribution of the phononic part of the bound state can be calculated
by solving the stationary Schr\"{o}dinger equation, yielding
\begin{eqnarray}\label{ME30}
C_{j,n}^{m}=\frac{gC_e}{2\pi}\int_{-\pi}^{\pi}dke^{ikj}\sum_{s}\frac{\hat{M}(n,s)\hat{M}^{-1}(s,m)}{E_{BS}-\omega_s(k)}.
\end{eqnarray}
Here, $C_{j,n}^{m}$ is the wavefunction at $n=a/b/c$ sublattice of the $j$ unit cell
when the spin is coupled to the $m=A/B/C$ sublattice at the $j=0$ unit cell.

Inspired by the fact that chiral spin-phonon bound states appear at the zero-energy modes in a dimer lattice,
we start to find the chiral bound states in a trimer lattice
by setting the energy of the bound states equal to that of the chiral phononic edge states.
For a finite system with hopping rates $(J_a, J_b,J_c)=(1,4,3)J$,
there exist two edge states localized at the right end with energies $\pm4J$.
Thus, we plot the phononic part of the bound states in Fig.~7(a,d)
with setting $E_\text{BS}=\pm4J$ in Eq.~(\ref{ME30}).
We find the spin-phonon bound state is chiral only when the spin is coupled to the $A$ sublattice,
i.e., $C_{j,n}^A=0$ for $j\geq0$ and $C_{j,n}^A\neq0$ for $j<0$.
Also, these bound states have no components in  the $A$ sublattice $C_{j,a}^A=0$,
while the amplitudes in the $B$ and $C$ sublattices are equal $|C_{j,b}^A|=|C_{j,c}^A|$.
Furthermore, the bound states are fractionally chiral when the spin is coupled to the $B$ and $C$ sublattices,
which is reflected in the components on the $A$ sublattice, $C_{j,a}^{B/C}=0$ for $j\leq0$ and $C_{j,a}^{B/C}\neq0$ for $j>0$.

However, the edge modes change when varying the boundary of the trimer lattice.
When the finite system is constructed by $(J_a, J_b,J_c)=(3,1,4)J$,
there are two pairs of edge states localized at the boundary, with energies $\pm3J$ and $\pm J$.
While for the finite system constructed by $(J_a, J_b,J_c)=(4,3,1)J$, there is no edge mode.
The bulk properties of the two cases are the same as that of $(J_a, J_b,J_c)=(1,4,3)J$.
Thus, we turn to study the spin-phonon bound states at these frequencies.
In Fig.~7(b,e) and 7(c,f), we show two pairs of chiral bound states located on the right/left side of the spin
with $E_\text{BS}=\pm3J/\pm J$, when the spin is coupled to the $B/C$ sublattice.
Note that the six chiral bound states are one-to-one corresponding to the six edge states.

We now consider the robustness of the chiral bound states in the trimer lattice.
As an example, we add disorder to the finite system with $20$ unit cells,
by setting $(J_a, J_b,J_c)=(1,4,3)J$, where a spin is coupled to the $B$ sublattice with energy $\omega_\sigma=\omega_m+3J$.
In Fig.~7(g), we allow disorder in the intracell hopping $J_a$ and $J_b$ independently.
While in Fig.~7(h), disorder in the intercell hopping $J_c$ is plotted.
On the other hand, bound states with on-site disorder acting in $\omega_m$ are shown in Fig.~7(i).
We observe several features from the numerical results.
First, the directionality is robust to disorder in the intracell hopping.
Second, the bound state has weight in each sublattices and both sides while the eigenenergies have a small deviation,
in the presence of disorder in the intercell hopping or on-site disorder.
Generally, the directional bound state is still robust when the disorder strength is on the order of  the coupling/hopping strength.
All these results  indicate that the directional spin-phonon bound states
can inherit the robustness of chiral edge states in the trimer lattice.

\begin{figure*}
\includegraphics[scale=0.22]{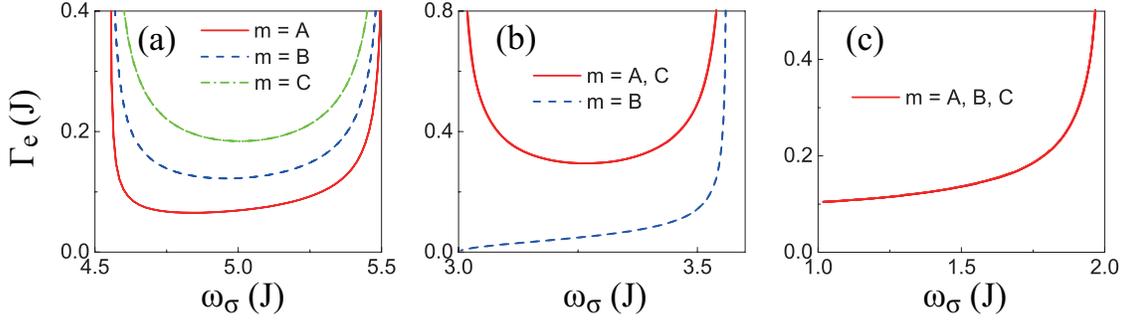}
\caption{\label{fig1}(Color online)
The decay rate $\Gamma_e$ of single spins coupled to $m=A/B/C$ sublattice in trimeried lattice.
(a) Inversion-symmetry broken phase $(J_a,J_b,J_c)=(1,4,3)J$.
(b) Inversion-symmetric phase $(J_a,J_b,J_c)=(1,1,3)J$. (c) $J_a=J_b=J_c=J$.}
\end{figure*}

\subsection{Sublattice-dependent spin relaxation}

From the discussion of chiral spin-phonon bound states in both lattices,
we can conclude that the spin indeed acts as
an effective boundary of the two semi-infinite waveguide structures.
In this case, the system can be divided into three parts:
the spin and two semi-infinite phononic waveguides.
Contrary to what happens in the dimer lattice
where the two semi-infinite structures are the same for  spin at the $A/B$ sublattice,
there are six different semi-infinite structures in the trimer lattice
when the spin is respectively coupled to the $A/B/C$ sublattice.
Thus, we predict that $\Sigma_e^A\neq\Sigma_e^B\neq\Sigma_e^C$
in the inversion-symmetry breaking phase of the trimer lattice,
while $\Sigma_e^A=\Sigma_e^C\neq\Sigma_e^B$ in the inversion-symmetric phase of the trimer model.
We note that the real part of the self-energy is the energy shift, which is small compared to the resonance frequency.
Furthermore, the shift is zero when the spin's frequency lies within the band.
Therefore, we mainly focus on the spin relaxation when the frequency is resonant with the phononic waveguide.

Without loss of generality, we consider a single spin with frequency within the upper band,
which is coupled to the $m\!=\!A/B/C$ sublattice, respectively.
In the Markovian regime, the spin's decay rate $\Gamma_e^m(\omega_\sigma)=-2\text{Im}\Sigma_e^m$ can be written as
\begin{eqnarray}\label{ME31}
\Gamma_e^m(\omega_\sigma)=\frac{2g^2\hat{M}(m,0)\hat{M}^{-1}(0,m)}{v_g(\omega_\sigma)},
\end{eqnarray}
with $v_g(\omega_\sigma)=\partial\omega/\partial k|_{\omega=\omega_\sigma}$ the group velocity.
In Fig.~8, we numerically plot the decay rate of the spin coupled to different sublattices
when the spin's frequency lies within the upper band.
The hopping rates are $(J_a, J_b,J_c)=(1,4,3)J$, $(J_a, J_b,J_c)=(1,1,3)J$ and $J_a\!\!=\!\!J_b\!\!=\!\!J_c\!\!=\!\!J$
in Fig.~8(a), 8(b) and 8(c), respectively.
We find $\Gamma_e^A\!\neq\!\Gamma_e^B\!\neq\!\Gamma_e^C$ for $J_a\!\neq\!J_b\neq\!J_c$.
In the special case $J_a\!=\!J_b\!\neq\!J_c$ where inversion symmetry recovers, $\Gamma_e^A=\Gamma_e^C\neq\Gamma_e^B$.
Moreover, the standard 1D band-edge divergence of $\Gamma_e^B$ is canceled
in the inversion-symmetric phase of the trimer model,
when $\omega_\sigma$ approaches the lower band-edge of $\omega_\alpha(k)$.
For comparison, the situation that the spin interacts with a normal coupled cavity waveguide is shown in Fig.~8(c),
which predicts a decay rate $\Gamma_e=g^2/|\sin(k)|$.
In general, the spin relaxation in a trimer lattice is sublattice-dependent, which is consistent with the discussion above.

\subsection{Bound state-mediated spin-spin interactions}

\begin{figure}
\includegraphics[scale=0.18]{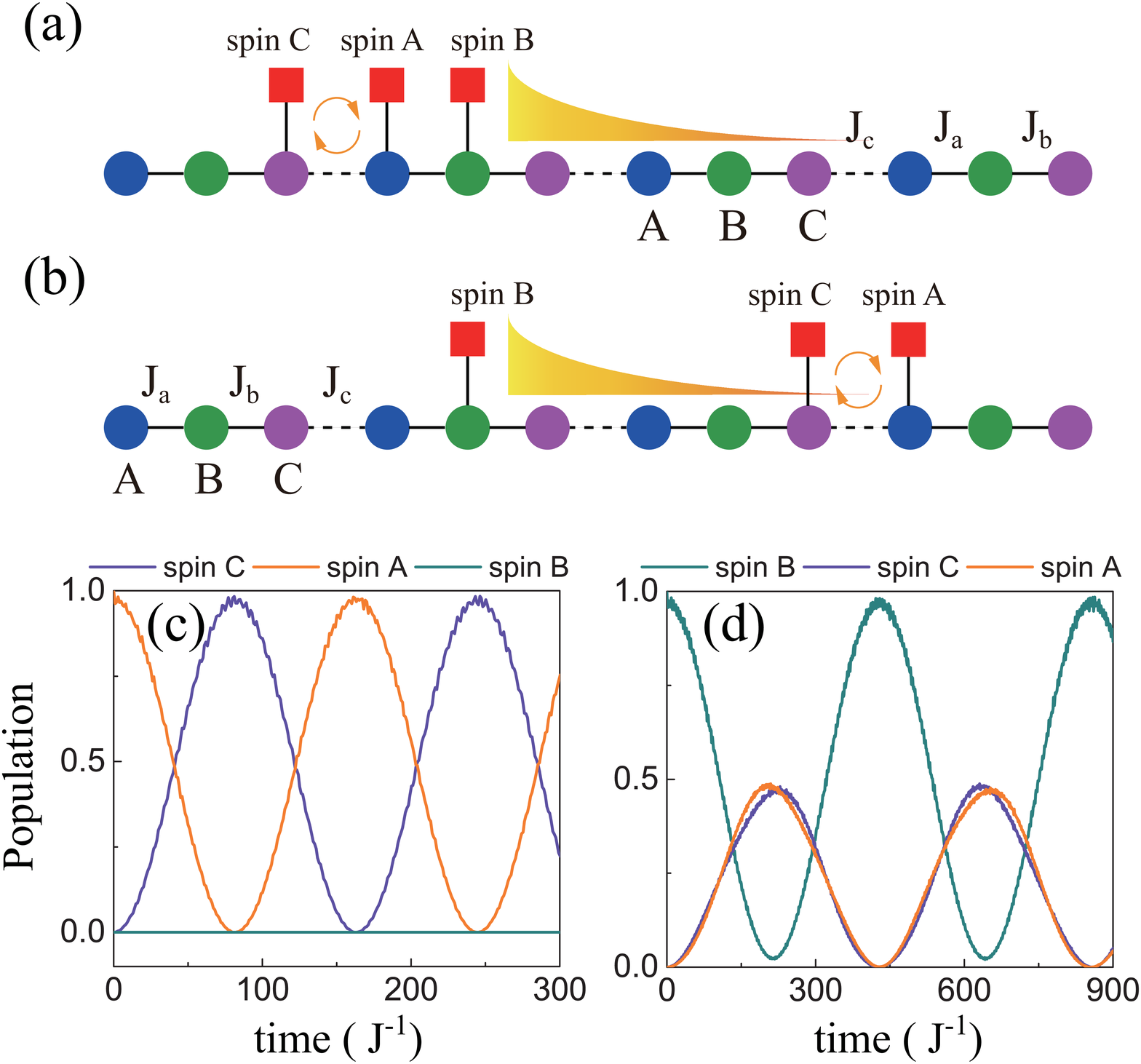}
\caption{\label{fig1}(Color online) Directional and relatively robust spin-spin interactions in trimer lattices.
(a,b) Three spins are coupled to a lattice chain with $12$ phononic cavities.
In (a), spins' position $(x_C,x_A,x_B)=(3,4,5)$. In (b), $(x_B,x_C,x_A)=(5,9,10)$.
(c,d) Time evolution of the population of three spins for the cases in (a) and (b), respectively.
The initial state in (c), $|\psi_0\rangle=|e\rangle_A|g\rangle_B|g\rangle_C$
and in (d), $|\psi_0\rangle=|g\rangle_A|e\rangle_B|g\rangle_C$.
The disorder in $J_a$ and $J_b$ is added independently with the same as that in Fig.~7(g).}
\end{figure}

We now discuss the spin-spin interactions mediated by
the chiral spin-phonon bound states appearing in the trimer lattice.
In the weak coupling limit, the interaction strength between two spins is
\begin{eqnarray}\label{ME32}
J_{ij}^{mn}=\frac{g^2}{2\pi}\int_{-\pi}^{\pi}dke^{ik(j-i)}\sum_{s}\frac{\hat{M}(n,s)\hat{M}^{-1}(s,m)}{E_{BS}-\omega_s(k)},
\end{eqnarray}
with $i,j$ the cell index, and $m,n$ the sublattice index.
As an example, we take advantage of the bound state shown in Fig.~7(b)   involving three spins
with one at the $B$ sublattice and two neighbouring spins at the $A/C$ sublattices.
When the spins at the $A/C$ sublattices are placed on the left side of the spin $B$, as shown in Fig.~9(a),
the two neighbouring spins are decoupled to the spin $B$ due to the directionality of the bound state and
the population can be transferred between the two neighbouring spins.
For the case of the spins at the $A/C$ sublattices on the right side of the spin $B$, as shown in Fig.~9(b), the excitation is exchanged
between the spin at the $B$ sublattice and the antisymmetry superposition of the two neighbouring spins.
The dynamics of the three spins can be modeled with an effective Hamiltonian in the interaction picture as
\begin{eqnarray}\label{ME33}
\hat{H}_\text{eff}&=&J_{B}(\hat\sigma_+^B\hat\sigma_-^C-\hat\sigma_+^B\hat\sigma_-^A+\text{H.c.})
+J_{AC}(\hat\sigma_+^A\hat\sigma_-^C+\text{H.c.})\notag\\
&=&\sqrt{2}J_{B}(\hat\sigma_+^B\hat\sigma_-^\text{at}+\text{H.c.})
+J_{AC}(\hat\sigma_+^\text{st}\hat\sigma_-^\text{st}-\hat\sigma_+^\text{at}\hat\sigma_-^\text{at}),
\end{eqnarray}
with $\hat{\sigma}_-^\text{st/at}=(\hat\sigma_-^C\pm\hat\sigma_-^A)/\sqrt{2}$ and $J_B=J_{ij}^{BC}=J_{ij+1}^{BC}$.
In the first situation $J_B=0$ for $j\geq i$, while in the second situation $J_B\neq0$ for $j<i$.
This reflects the directionality of the spin-spin interaction when the spin is coupled to the $B$ sublattice.
When focusing on the case $J_B\neq0$,
we find that the transition is not resonant due to the interaction between the neighbouring spins.
Note that the detuning (the last term in the second line of Eq. (\ref{ME33})) can be removed by external driving fields;
thus the antisymmetric combinations of spin $A$ and spin $C$
acting as a single spin can be resonantly coupled to spin $B$.
Finally, we emphasize that the spin-spin interaction is directional only when one of the spins is at the $B$ sublattice.

We consider a  resonator chain composed of $12$ lattice sites to numerically examine the two cases,
as shown in Fig.~9. The disorder in the intracell hopping rate $J_a$ and $J_b$ is also taken into account
and the spin's frequency is $\omega_\sigma\simeq3J$ in both cases.
In the first case, the three spins are placed at the third, $4$th and $5$th lattice sites
and the Rabi oscillation between spin $A$ and spin $C$ is observed in Fig.~9(c),
with the initial state $|\psi_0\rangle=|e\rangle_A|g\rangle_B|g\rangle_C$.
We show the population in spin $B$ keeps zero.
In the second case, the three spins are placed at the $5$th, $9$th and $10$th lattice sites
and the population is transferred between spin $B$ and spin $A/C$
with the initial state $|\psi_0\rangle=|g\rangle_A|e\rangle_B|g\rangle_C$, as shown in Fig.~9(d).
Though the energy of the bound state formed by spin $B$ is strictly pinned under disorders ($E_\text{BS}^B=3J$),
the energy of the bound states formed by spin $A$ and spin $C$ are not strictly pinned ($E_\text{BS}^{A/C}\approx3J$).
Therefore, the population of the spin $B$ can not reach zero with the time evolution.

\subsection{Quantum control over chiral phononic edge states}

\begin{figure}
\includegraphics[scale=0.23]{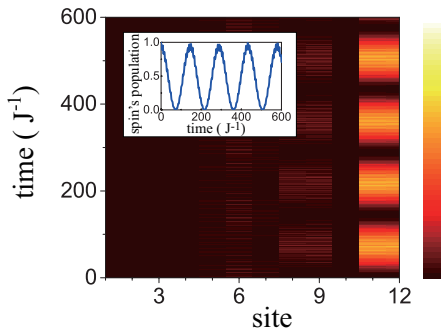}
\caption{\label{fig1}(Color online)
Control over the phononic edge states in the trimer lattice
composed of $12$ phononic cavities, with a single spin inside the $5$th cavity.
Time evolution of the population in the cavity array and the spin is plotted.
Here, we choose $(J_a,J_b,J_c)=(1,4,3)J$.}
\end{figure}

Because the edge modes in the trimer lattice remain in the bandgap,
the quantum control on these modes becomes possible with the addition of spins.
Furthermore, since the edge states in the inversion-symmetry  breaking phase
of the trimer lattice have no degeneracy,
we can directly manipulate one of the phononic edge states
on the boundary of the phononic crystal with the addition of a single spin.
The resonance transfer can occur via tuning the frequency of the spin, with the interaction Hamiltonian
\begin{eqnarray}\label{ME34}
\hat{H}_{e-s}=g_{e-s}(|E\rangle\langle e|+|e\rangle\langle E|),
\end{eqnarray}
where $g_{e-s}\!=\!g\langle E|\hat{d}_{x_0}^\dag|\text{vac}\rangle$ is the coupling rate.
As an example, we consider the situation that a single spin is placed at the $5$th cavity of a $12$ cavity chain,
with alternating hopping strengths $(J_a,J_b,J_c)=(1,4,3)J$.
There exist two edge states localized on the right side of the system with energy $E_\text{edge}=\pm4J$.
In Fig.~10, we plot the time evolution of the system and show a transition
with the amplitude approaching one between the spin and one of the chiral phononic edge states.

\section{experimental feasibility}

In this section, we discuss the experimental feasibility of our system.
As illustrated in Fig.~1(a), an array of defect cavities can be fabricated in a diamond nanobeam.
As an example, we consider the phononic cavities with resonance frequency $\omega_m/2\pi\sim5$ GHz
and two adjacent phononic cavities are coupled via a phononic waveguide.
In the rotating frame, the cavity-waveguide-cavity Hamiltonian takes the form of
\begin{eqnarray}\label{ME35}
\hat{H}_\text{cwc}&=&\sum_{k}\delta_k\hat{f}_k^\dag\hat{f}_k\notag\\
&+&\sum_{k}\{(g_{k,1}\hat{d}_1+g_{k,2}(-1)^k\hat{d}_2)\hat{f}_k^\dag+\text{H.c.}\},
\end{eqnarray}
where $\delta_k$ is the detuning between the cavity and the $k$th waveguide mode,
$\hat{f}_k^\dag$ ($\hat{f}_k$) is the creation (annihilation) operator of the $k$th waveguide mode,
$g_{k,i}$ is the cavity-waveguide coupling and the term $(-1)^k$ accounts for the symmetry of the waveguide modes.
Consider that the waveguide is short and a single mode is nearly resonant with the cavity,
the Hamiltonian can be simplified to
\begin{eqnarray}\label{ME36}
\hat{H}_\text{cwc}&=&\delta\hat{f}^\dag\hat{f}
+g_1(\hat{d}_1\hat{f}^\dag+\hat{f}\hat{d}_1^\dag)\notag\\
&+&g_2(\hat{d}_2\hat{f}^\dag+\hat{f}\hat{d}_2^\dag).
\end{eqnarray}
When taking the large detuning limit $g_i\ll\delta$ and applying the perturbation theory,
we can arrive at the effective Hamiltonian describing hopping interactions between the two adjacent cavities
\begin{eqnarray}\label{ME37}
\hat{H}_\text{cc}=J_\text{hop}(\hat{d}_1^\dag\hat{d}_2+\hat{d}_2^\dag\hat{d}_1),
\end{eqnarray}
with $J_\text{hop}=g_1g_2/\delta$ the effective hopping rate between the phononic cavities.
The cavity-waveguide coupling is tunable via designing the mirror section of the nanocavity,
which connects the cavity and the waveguide.
Meanwhile, the detuning can be tuned via changing the structure of the connecting waveguide.
So far, the hopping strengths between adjacent mechanical resonators
have been designed and realized with a few megahertz
\cite{Generalized2017Fang,PhysRevX.8.041027}.

We now show how to design the lattice structure in this work.
As an example, we consider the case of the dimer lattice.
The phononic cavities are designed with different mirror sections on the left and right sides:
one with a large number of phononic crystal periods and the other with a smaller number of periods,
forming a strong mirror and a weak mirror respectively.
The strong/weak mirror can give rise to small/large cavity-waveguide coupling
\cite{PhysRevX.8.041027,Optical2016Fang,Generalized2017Fang,Mirhosseini2020Mirhosseini}.
Thus, if we always align the strong mirror section with the strong mirror section
and use identical waveguide to connect the nanocavities,
the hopping strengths are alternating with the strengths $J_1=g_w^2/\delta$ and $J_2=g_s^2/\delta$.
Note that the trimer lattice can be engineered with the same method.

For a typical nanobeam cavity shown in Refs.\!
\cite{Optical2016Fang,Generalized2017Fang,Mirhosseini2020Mirhosseini},
it has an intrinsic decay rate $\gamma_i$ on the order of MHz.
However, the phonons are virtually excited in the process of spin-spin interactions,
where the excited number of the phonons $\sim0.01$ is at a low level.
Furthermore, we can use phononic shielding to surround the nanobeam (the whole network),
which can provide a phononic bandgap to mitigate the leakage to the bulk
\cite{PhysRevX.8.041027}.
This protection of phonon modes can result in high quality factor $Q$ exceeding $10^7$,
and even ultra-high $Q\sim10^{10}$ of nanocavities in the 1D/2D optomechanical crystal
has been demonstrated in  recent experiments
\cite{MacCabe840,Two2020Ren}.
Correspondingly, the intrinsic damping can be reduced to below kilohertz ($\gamma_i/2\pi<1$ kHz).

SiV centers are point defects in diamond with a silicon atom occupying two adjacent vacancies,
which can be embedded into diamond crystal via ion implantation technology.
As depicted in Fig.~1(c), the phononic modes can be directly coupled to the spin in a single SiV
through matching the spin's frequency to the phonon frequency,
where an external magnetic field tilted from the symmetry axis of the defect should be applied.
The spin-phonon coupling strength can be estimated by \cite{PhysRevLett.120.213603,PhysRevB.97.205444}
\begin{eqnarray}\label{ME38}
g=\frac{d_\text{spin}}{v}\sqrt{\frac{\hbar\omega_m}{2\rho V}}\xi(\vec{r}_\text{SiV}).
\end{eqnarray}
Here, $d_\text{spin}/2\pi\sim100$ THz/strain is the strain sensitivity,
$v\sim10^4$ m/s is the group velocity of acoustic waves in diamond,
$\omega_m/2\pi\sim5$ GHz, $\rho\sim3500$ kg/$\text{m}^3$,
$V$ is the volume of the optomechanical nanocavity and $\xi(\vec{r}_\text{SiV})\sim1$
is the strain distribution at the position of the SiV centers.
Thus, for a typical nanocavity, the single spin-phonon coupling strength can reach $g/2\pi\sim1$ MHz.

The main limit of the coherence of SiV centers is the dephasing time $T^*_2$,
which arises from the coupling to the thermal environment and the nuclear spins in diamond.
The thermal occupation can be frozen out in the low temperature
and the interactions with the nuclear spins can be suppressed using the dynamical decoupling technology
\cite{Hanson352,Preserving2009Du,PhysRevB.92.224419}.
For the dilution refrigeration temperature $T\sim10$ mk,
the spin dephasing $\gamma_s/2\pi\sim100$ Hz has been realized in the experiment
\cite{PhysRevLett.119.223602}.
In general, we choose $J/2\pi\sim3$ MHz and $g/2\pi\sim1$ MHz in the main text,
which leads to the spin-spin coupling strength $J_{ij}/2\pi\sim100$ kHz.
The dissipations of the spins and the phononic crystal can be neglected,
since $\{\gamma_i,\gamma_s\}\ll J_{ij}$.

\section{conclusion}

We have proposed a feasible scheme in the solid-state platform to
explore novel interactions between spins and phonons in dimer and trimer lattices,
with an array of SiV centers integrating into a 1D optomechanical crystal.
We have predicted a series of unconventional phenomena in this spin-phononic hybrid system.

In the dimer lattice, we show the formation of chiral
and robust spin-phonon bound states, which can mediate directional and robust spin-spin interactions.
We also show topology-dependent phononic collective radiation effects.
The coupling between a single spin and the phononic edge modes is discussed as well.
While in the trimer lattice, we show that the chiral and robust spin-phonon bound states still exist and the spin relaxation is sublattice-dependent.
The bound state-mediated spin-spin interactions exhibit chiral features and are relatively robust.
Quantum control over the chiral edge states, which are unique in the inversion-symmetry breaking phase of the trimer lattice,
is also demonstrated with the addition of spins.

Finally, we discuss the feasibility of this scheme under realistic parameters.
This work takes the advantage of phononic crystals and SiV centers,
thus providing a promising platform for manipulating phonons with spins,
and is useful for phonon-mediated quantum information   applications.

\section*{Acknowledgments}
We gratefully acknowledge the use of the open source
Python numerical packages QuTiP
\cite{JOHANSSON20121760,JOHANSSON20131234}.
This work is
supported by the National Natural Science Foundation of
China under Grant No. 92065105  and
Natural Science Basic Research Program of Shaanxi
(Program No. 2020JC-02).


\bibliography{ref}

\end{document}